%% file: final.tex
\documentclass[journal,twocolumn]{IEEEtran}
\usepackage{epsfig,makeidx,color}
\usepackage{amsmath,amssymb,amsthm,bbm}
\usepackage{cite,graphicx}
\usepackage{enumerate}
\usepackage{hyperref}

\def\defh{
\mbox{\footnotesize $ \begin{array}{c} \cH_1 \cr > \cr < \cr \cH_0 \end{array} $}}

\def\cV{{\cal V}}

\def\cH{{\cal H}}

\def\cX{{\cal X}}

\def\cQ{{\cal Q}}

\def\rH{{\rm H}}

\def\rT{{\rm T}}

\def\uC{{\mathbb C}}
\def\uP{{\mathbb P}}
\def\uE{{\mathbb E}}

\DeclareMathOperator*{\argmin}{\arg\!\min}
\DeclareMathOperator*{\argmax}{\arg\!\max}

\newtheorem{mytheorem}{\bf Theorem} 

\def\deft{ \buildrel \triangle \over = }

\def\be{ \begin{equation} }
\def\ee{ \end{equation} }
\def\bea{ \begin{eqnarray} }
\def\eea{ \end{eqnarray} }
\def\bx{{\bf x}}
\def\by{{\bf y}}
\def\bc{{\bf c}}

\def\bb{{\bf b}}

\def\bg{{\bf g}}

\def\ba{{\bf a}}

\def\bn{{\bf n}}

\def\bA{{\bf A}}

\def\bG{{\bf G}}
\def\bH{{\bf H}}
\def\bI{{\bf I}}

\def\bV{{\bf V}}

\def\bR{{\bf R}}

\def\bX{{\bf X}}

\def\bone{{\bf 1}}

\def\b0{{\bf 0}}

\def\cC{{\cal C}}
\def\cD{{\cal D}}

\def\cN{{\cal N}}
\def\cS{{\cal S}}

\ifCLASSOPTIONonecolumn
  \newcommand{\figwidth}{0.60\columnwidth}
\else
  \newcommand{\figwidth}{0.95\columnwidth}
\fi

\begin{document}

\title{Matched-Filter based
Backscatter Communication for IoT Devices over Ambient OFDM Carrier}

\author{Jinho Choi\\
\thanks{The author is with
School of Information Technology,
Deakin University, Geelong, VIC 3220, Australia.
(e-mail: jinho.choi@deakin.edu.au).}}

\date{today}
\maketitle

\begin{abstract}
In this paper, we study backscatter communication (BC)
for power-limited devices that are connected to a network
for the Internet of Things (IoT),
where joint estimation and detection 
is carried out at a receiver to detect signals from 
a backscatter device (BD) 
with ambient orthogonal frequency division multiplexing (OFDM) carrier.
In conventional BC, in 
order to avoid the difficulty of the carrier estimation,
the energy detector is usually considered at 
a receiver at the cost of poor performance.
To improve the performance, in this paper,
we consider a novel approach 
that allows the carrier estimation at the receiver
via joint estimation and detection.
In particular, in the proposed approach,
the matched-filtering 
at the BD (for the transmitter filter) 
is employed to impose a certain
property that allows efficient and reliable
carrier estimation via joint estimation and detection.
Through the performance analysis and simulation results,
we show that the matched-filtering 
at the BD in the proposed approach
can improve the performance.

\end{abstract}

{\IEEEkeywords
Internet of Things;
backscatter communication; OFDM; carrier estimation}

\ifCLASSOPTIONonecolumn
\baselineskip 26pt
\fi

\section{Introduction} \label{S:Intro}

The Internet of Things (IoT) 
has attracted attention from both academia and practitioners 
and is considered for standardizations \cite{ITU_IoT} \cite{Keoh14}.
The IoT is to connect any device to the Internet through wired or wireless
communications for various applications and services.
While 
some devices might be quite capable,
e.g., smart phones, most cheap IoT
devices would have various limitations.
In particular, some IoT devices may rely on
the power from radio frequency (RF) signals, i.e., RF powered devices,
to compute and communicate with other devices \cite{Gollakota14}.
In this case, backscatter communication would play a crucial role
in transmitting information to nearby receivers such as RF identification
(RFID) systems \cite{Dobkin12} \cite{Boyer14}.
There are a number applications
of RFID technology. For example,
in \cite{Yu19} and \cite{Luo19}, RFID is used
for inventory and localization, respectively.
In addition, in \cite{Ma18}, for networking with miniature medical
sensors or devices implanted in deep tissues, RFID technology
is considered.

While RFID requires dedicated RF transmitters,
it is also possible to exploit other existing or ambient RF signals 
\cite{Liu13}.
Ambient backscatter communication (AmBC) has been 
studied for devices of limited power (such as sensors),
which are called backscatter devices (BDs), in the IoT 
to transmit their information to nearby receivers
by exploiting various ambient RF signals
\cite{Liu13} \cite{Kellogg14} \cite{Wang16}. 
For the source of ambient RF signals,
TV and WiFi signals can be considered \cite{Liu13} \cite{Bharadia15}.
In \cite{Kang17}, 
a ambient backscatter technique using multiple antennas is studied
to achieve a high data rate in AmBC.
In \cite{Yang18B}, a cooperative AmBC approach is 
investigated where a reader recovers information not only from a
BD, but also from an RF source.
In \cite{Yang19}, a full-duplex based
AmBC network is proposed where an access point 
(AP) can simultaneously transmits downlink 
OFDM signals to its user and receives uplink signals backscattered 
from multiple BDs.

In AmBC, there are various challenges such as 
direct-link interference.
Unlike RFID systems,
an ambient signal is not a tone, but a modulated signal.
Provided that the ambient signal is a narrowband signal,
in \cite{Wang16}, differential modulation scheme
is employed for backscattered signals so that the energy
of direct-link interference can be mitigated.
If the ambient signal is not a narrowband signal 
(e.g., WiFi signals) and
experiences frequency-selective fading, the mitigation of 
\emph{wideband} direct-link interference becomes more challenging.
In \cite{Bharadia15}, the notion of BackFi
is introduced where a WiFi AP is not only 
a reader (or receiver for backscattered signals),
but also an ambient transmitter. Since the AP transmits signals
(which are ambient signals for a BD), it 
can employ self-interference-cancellation using reflected signals.
However, if the ambient transmitter and the receiver are different,
the mitigation of wideband direct-link interference
becomes hard as the receiver cannot directly observe
the channel state information (CSI)
from the ambient transmitter to a BD. 
To avoid this problem,
in \cite{Yang18}, a special structure of 
orthogonal frequency division multiplexing (OFDM) signals,
which is called the repeating structure, is exploited 
to suppress wideband direct-link interference 
in detecting backscattered signals.

Ambient OFDM carrier \cite{Yang18} is an attractive approach
for backscatter communication
as OFDM systems are popular, e.g., 
long-term evolution (LTE) systems for cellular 
communications \cite{Dahlman13}, WiFi, and 
digital video broadcasting - terrestrial (DVB-T)
\cite{Eizmendi14}.
Thus, in this paper, we consider ambient OFDM carrier.
However, unlike \cite{Yang18}, we do not rely on
the repeating structure of OFDM signals.
The reasons are two-fold: {\it i)} The repeating
structure is due to cyclic prefix (CP). 
If the length of CP is not sufficiently long,
it is not possible to exploit the repeating structure.
{\it ii)} When the repeating structure is used 
to detect backscattered signals, 
the performance is independent of the length of OFDM symbols.
However, in general, 
a better performance is expected if the length of OFDM symbols increases
as the bit energy of backscattered signals increases.

In this paper, 
we consider coherent detection of backscattered signals
via joint estimation and detection for a better
detection performance in AmBC.
Note that although coherent detection can provide a better performance
than noncoherent detection 
(which is widely adopted to detect backscattered signals) 
\cite{ProakisBook},
it has more challenges including the carrier estimation.
For the carrier estimation, the receiver needs
to know BD's CSI, which is not directly observable at the receiver.
As a result, in conventional approaches,
reliable carrier estimation becomes
difficult as the receiver cannot directly estimate BD's CSI.
To address this challenge,
we propose the matched-filtering at the BD (for the transmitter filter)
to impose a certain
property that allows efficient and reliable
carrier estimation via joint estimation and detection
at the receiver. Furthermore,
since joint estimation and detection at the receiver usually
requires a high-computational complexity,
we consider an iterative method based on the expectation-maximization
(EM) algorithm \cite{Dempster77} \cite{EM_Book},
where the complexity of each iteration is linearly proportional
to the length of OFDM symbol.
The number of iterations to convergence is also a few
(at a high signal-to-noise
ratio (SNR), 5 iterations are shown to be sufficient).

The work in this paper differs from other existing works.
Unlike \cite{Bharadia15}, it does not require
that the ambient transmitter is also the receiver
to mitigate wideband direct-link interference.
Furthermore, as opposed to \cite{Yang18},
the proposed approach is applicable even if the length of CP is short.
since the repeating structure is not employed.
In \cite{Qian17}, as in this paper,
the estimation of CSI is 
considered for semi-coherent detection.
However, only flat fading channels are considered in 
\cite{Qian17}, while frequency-selective
fading channels are assumed in this paper. 
In addition, we consider full coherent detection.

The rest of the paper is organized as follows.
In Section~\ref{S:SM}, we present an OFDM-based AmBC model 
and discuss difficulties due to
wideband direct-link interference.
Existing approaches to detect
backscattered signals with ambient OFDM carrier
are presented in Section~\ref{S:EA}.
We propose a new approach that effectively allows
coherent detection of scattered signals by
taking advantage of known CSI 
and present a low-complexity joint 
estimation and detection method
in Section~\ref{S:MAT}.
A performance analysis is presented in Section~\ref{S:PA}.
In Section~\ref{S:Sim}, we present simulation results.
The paper is concluded with some remarks in Section~\ref{S:Conc}.

{\it Notation}: 
Matrices and vectors are denoted by upper- and lower-case
boldface letters, respectively.
The superscripts $\rT$ and $\rH$
denote the transpose and complex conjugate, respectively.
The $2$-norm of a vector $\ba$ is denoted by $|| \ba ||$.
The Frobenius norm of a matrix $\bA$ is denoted by $||\bA||_{\rm F}$.
$\uE[\cdot]$
and ${\rm Var}(\cdot)$
denote the statistical expectation and variance, respectively.
$\cC \cN(\ba, \bR)$
represents the distribution of
circularly symmetric complex Gaussian (CSCG)
random vectors with mean vector $\ba$ and
covariance matrix $\bR$.
The Q-function is given by
$\cQ(x) = \int_x^\infty \frac{1}{\sqrt{2 \pi} } e^{- \frac{t^2}{2} } dt$.

\section{System Model} \label{S:SM}

\begin{figure}[thb]
\begin{center}
\includegraphics[width=\figwidth]{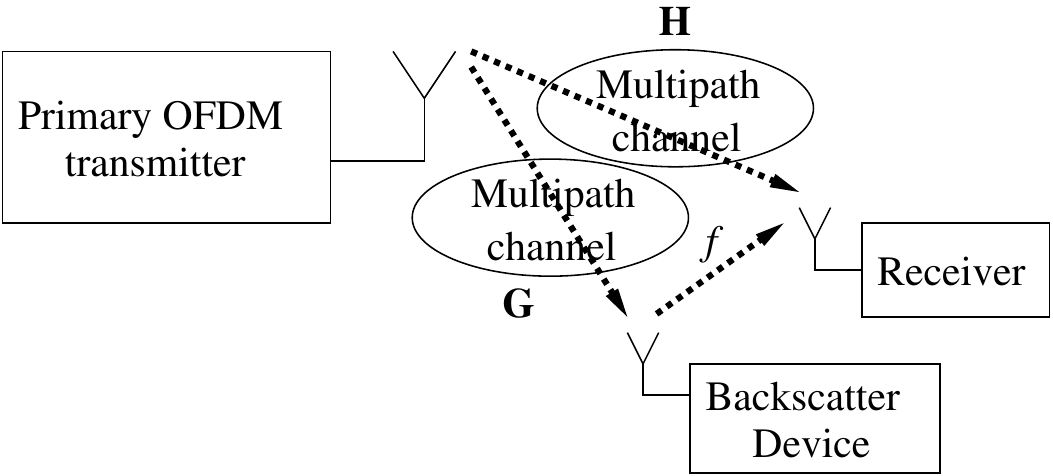}
\end{center}
\caption{An illustration of the system consisting of
the legacy OFDM transmitter, BD, and receiver. Here,
$\bH$ and $\bG$ represent
the frequency-domain 
(wideband) channel matrices from the legacy OFDM transmitter
to the receiver and the BD, respectively,
while $f$ represents the channel coefficient
from the BD to the receiver.}
        \label{Fig:sys}
\end{figure}

Consider a system consisting of a 
legacy (or primary) OFDM transmitter
as an RF source (i.e., an ambient transmitter), 
a BD, and a receiver that is a reader for the BD as
in \cite{Yang18}.
The system is illustrated as in Fig.~\ref{Fig:sys}.
As discussed in \cite{Bharadia15},
there are IoT gadgets such
as cheap wearable devices and sensors that can measure physical variables 
to upload to the cloud through a gateway.
These gadgets and devices are to operate for a long time
without requiring battery replacements or without batteries.
To this end, exploiting ambient signals such as WiFi signals from 
nearby APs to transmit their information becomes crucial.
According to the above scenario, 
in Fig.~\ref{Fig:sys},
the primary OFDM transmitter, the BD, 
and the receiver are
a WiFi AP, a cheap IoT gadget, and a gateway, respectively.
Note that as shown in Fig.~\ref{Fig:sys},
the receiver is not able to observe or estimate BD's CSI, which 
is denoted by $\bG$,
while it can estimate its CSI, which is denoted by $\bH$.
We will explain $\bH$ and $\bG$ later.

For convenience, the system consisting of a legacy OFDM transmitter
and legacy receivers is referred to as the legacy system.
Denote by $\{h_p\}$ the channel
impulse response (CIR) from the legacy OFDM transmitter
to the receiver and by $\{g_p\}$ the CIR 
the legacy OFDM transmitter to the BD.
Since the bandwidth of the legacy OFDM system
is wide, the lengths of CIRs are relatively long and
the channels experience intersymbol-interference (ISI) and
become frequency-selective \cite{ChoBook, ProakisBook}.
Let $\bx_m$
denote the $m$th (ambient) OFDM symbol transmitted
by the legacy OFDM transmitter. We assume that the length 
of $\bx_m$ is $L$, 
i.e., $\bx_m = [x_{0,m} \ \ldots \ x_{L-1,m}]^\rT$,
where $x_{l,m}$ represents the signal transmitted through
the $l$th subcarrier.
Throughout the paper, we assume that
$$
\uE[x_{l,m}] = 0, {\rm Var}(x_{l,m}) = E_x, 
\ \mbox{and} \ x_{l,m} \in \cX,
$$
where $\cX$ is the signal constellation of $x_{l,m}$.
In order to deal with the 
ISI, CP is added \cite{ChoBook} \cite{ChoiJBook}.
At the receiver, the received signal 
after removing CP (in the frequency-domain) becomes
\be
\by_m = \bH \bx_m + \ba_m + \bn_m, \ m = 0, \ldots, M-1,
	\label{EQ:bym}
\ee
where $\bH = {\rm diag}(H_0, \ldots, H_{L-1})$,
$\ba_m$ denotes the received signal from the BD
(i.e., the backscattered signal),
$\bn_m \sim \cC \cN(0, N_0 \bI)$ is the background noise vector,
and $M$ represents the number of OFDM symbols in a slot.
Here, 
we assume that the length of slot is shorter than
the coherence time so that the channel variation over the 
time is negligible and
$H_l$ is the channel coefficient of the $l$th subcarrier,
which is given by
$$
H_l = \sum_{p=0}^{P_h-1} h_p e^{- \frac{j 2 \pi lp}{L}},
\ l = 0,\ldots,L-1,
$$
where $P_h$ represents the length of the CIR from the legacy
OFDM transmitter to the receiver.
Similarly, the channel coefficient of the $l$th subcarrier
of the channel from the legacy OFDM transmitter to the BD is given by
$$
G_l = \sum_{p=0}^{P_g-1} g_p e^{- \frac{j 2 \pi lp}{L}},
\ l = 0,\ldots,L-1,
$$
where $P_g$ represents the length of the CIR from the legacy
OFDM transmitter to the BD.

We can assume that the receiver is sufficiently close to the BD
and the channel from the BD to the receiver can be modeled as
a flat fading channel \cite{Yang18} \cite{Bharadia15},
while the channels from the legacy OFDM transmitters
to the BD and receiver are modeled as frequency-selective
fading channels as shown above (due to multiple paths).
Consequently, although the system bandwidth is wide,
the channel coefficients for all the
subcarriers from the BD to the receiver are the same and denoted
by $f_l = f$, $l = 0, \ldots, L-1$. We can have
\be
\ba_m = \beta \bG \bX_m \bb_m,
	\label{EQ:bzm}
\ee
where $\beta = \alpha f$, $\bG = {\rm diag}(G_0, \ldots, G_{L-1})$,
$\bX_m = {\rm diag}(\bx_m)$,
and $\bb_m$ is the Fourier transform of the BD's baseband signal 
to be transmitted to the receiver.
Here, $\alpha$ is the reflection coefficient that is usually small
\cite{Dobkin12}.
In \eqref{EQ:bzm}, the 
diagonal elements of $\bG \bX_m$ (i.e.,
$\bG \bx_m$) is the received signal in the
frequency-domain at the BD,
which is regarded as the Fourier transform of the baseband representation
of the carrier that is to be exploited by the BD for 
backscatter communication.
For convenience, denote by $\tilde \bb_m$ 
and $\tilde \bc_m$ the time-domain representations of $\bb_m$
and $\bG \bx_m$, respectively.
Then, the time-domain representation of
$\ba_m$ is given by
$$
\tilde \ba_m = \beta \tilde \bc_m  \otimes \tilde \bb_m,
$$
where $\otimes$ represents the convolution.
The modulation in \eqref{EQ:bzm} is called
``modulation in the air" in \cite{Yang18},
where $\bG \bx_m$ or $\bG \bX_m$ becomes 
the carrier for backscatter communication.
Then, the received
signal, $\by_m$, in \eqref{EQ:bym} is rewritten as
\begin{align}
\by_m 
& = \bH \bx_m + \ba_m + \bn_m \cr
& = \bH \bx_m + \beta \bG \bX_m \bb_m
+ \bn_m, \ m = 0, \ldots, M-1.
	\label{EQ:bym1}
\end{align}

We have some remarks as follows.
\begin{itemize}
\item With ambient OFDM carrier, as shown in \eqref{EQ:bym1},
there are two main difficulties to detect backscattered signals. 
The first difficulty, denoted by
{\bf Difficulty-I}, is due to unknown
carrier, $\bH \bx_m$.
For the \emph{carrier estimation} to address the first difficulty,
the receiver can exploit pilot signals 
from the legacy transmitter, which are 
transmitted prior to transmissions of OFDM (data) symbols so that 
legacy receivers are able to estimate CSI.
Thus, the receiver can easily estimate $\bH$.
However, it is still required 
to estimate $\bx_m$ for the carrier estimation. 
The second difficulty, denoted by
{\bf Difficulty-II}, is due to unknown BD's CSI, $\bG$,
which is not directly observable at the receiver.


\item The CIRs, $\{h_p\}$ and $\{g_p\}$, include
the propagation delays. Thus, the first few coefficients can be zero.
In addition, the length of CP should be greater than 
or equal to the length of CIR, $P_h$,
to avoid any ISI between adjacent OFDM symbols.
However, it is also expected to make the length of CP as short as possible
to maximize the spectral efficiency. Thus, the length of CP might be
the same as the (maximum) length of CIR.

\item Since there might be the background or thermal noise at the BD,
the transmitted signal from the BD needs to include
the background noise. However, the thermal noise is negligible
as the BD might have passive components and limited signal processing
operations \cite{Wang16}.
\end{itemize}

\section{Existing Approaches}	\label{S:EA}

As mentioned earlier, the carrier estimation might be necessary
to detect backscattered signals in AmBC.
However, there are few existing approaches to avoid the carrier 
estimation. We briefly explain them in this section together with
their shortcomings.

\subsection{With Repeating Structure}

In this subsection, we briefly discuss the approach
studied in \cite{Yang18} and explain why this approach cannot
be used if there is no
additional guard interval between OFDM symbols in a slot.

In \cite{Yang18}, 
a transceiver design is considered by exploiting 
certain properties of OFDM signals with CP.
For simplicity, let $M = 1$ and
omit the OFDM symbol index $m$ in a slot.
Consider the received signal in the time-domain
at the receiver without backscattered signals as follows:
$$
\tilde y_t = \sum_{p=0}^{P_h-1} h_p \tilde x_{t-p} + \tilde n_t, \ 
t = 0, \ldots, L+ L_{\rm cp} - 1,
$$
where $L_{\rm cp}$ represents the length of CP
and $\tilde x_t$ and $\tilde n_t$ 
denote the $t$th signal and noise at the time-domain, respectively.
As CP is inserted, we have \cite{ChoiJBook} \cite{ChoBook}
$$
\tilde x_t = \tilde x_{L+t}, \ t \in \{0, \ldots, L_{\rm cp} - 1\}.
$$
The above repetition results in
the following property called the repeating structure  in \cite{Yang18}:
\be
\tilde y_t = \tilde y_{L+t}, \ t \in \{P_h, \ldots, 
L_{\rm cp} - 1\},
	\label{EQ:tyy}
\ee
which plays a key role in deriving the transceiver in \cite{Yang18}.

Let $\tilde b_t$ denote the signal transmitted by the BD in the 
time-domain. To exploit the repeating structure in \eqref{EQ:tyy}, 
the following signal is considered in \cite{Yang18}:
\begin{align}
\mbox{Bit 0}:\ & \tilde b_t = 1, \ t = \{0, \ldots, L+L_{\rm cp} - 1\} \cr
\mbox{Bit 1}:\ & \tilde b_t = \left\{
\begin{array}{rl}
1, & \mbox{if} \ t = \{0, \ldots, \bar L\} \cr
-1, & \mbox{if} \ t = \{\bar L, \ldots, L+L_{\rm cp} - 1\},\cr
\end{array}
\right.
\end{align}
where $\bar L = \lfloor \frac{L+L_{\rm cp} - 1}{2} \rfloor$.
With the signal transmitted from the BD, the 
received signal at the receiver in the time-domain becomes
\be
\tilde y_t = h_t \otimes \tilde x_t + \beta (g_t \otimes
\tilde x_t) \tilde b_t + \tilde n_t.
\ee
Let
\be
\tilde d_t = \tilde y_t - \tilde y_{L+t}, \ t \in 
\{P_h, \ldots, L_{\rm cp} - 1\}.
\ee
Then, if bit 0 is transmitted from the BD, 
we have
$$
\tilde d_t = \tilde e_t \deft \tilde n_t - \tilde n_{t + L}, 
\ t \in \{P_h, \ldots, L_{\rm cp} - 1\},
$$
where the signals from the legacy OFDM transmitter are canceled
thanks to \eqref{EQ:tyy}. On the other hand,
if bit 1 is transmitted, it follows
$$
\tilde d_t = 2 (g_t \otimes \tilde x_t) +
\tilde e_t, \ t \in \{P_h, \ldots, L_{\rm cp} - 1\},
$$
where the signal from the legacy OFDM transmitter exists.
Thus, the energy detector \cite{Scharf} can be used to detect
the signal from the BD as in \cite{Yang18}.

Unfortunately, 
since the approach in \cite{Yang18}
relies on the repeating structure
in \eqref{EQ:tyy}, there are few difficulties.
Clearly, it requires that $L_{\rm cp} > P_h$ and a better performance
can be achieved as $L_{\rm cp}$ increases, which results in an inefficient
use of bandwidth (or subcarriers) and undesirable as mentioned earlier.
Another difficulty is that
\eqref{EQ:tyy} is only valid when $M = 1$ or 
there are additional guard intervals
between OFDM symbols for the case
of $M > 1$. If $M > 1$, the received signals corresponding CP
include interference from the previous OFDM symbol, which
cannot guarantee \eqref{EQ:tyy} due to the repeating structure unless
the length of CP is sufficiently longer than the
length of CIR. Note that
since CP is used to avoid ISI, additional
guard intervals might be redundant to the legacy system
and would not be allowed.
Consequently, in this paper, we do not 
further consider the repeating structure.

\subsection{Without Repeating Structure}	\label{SS:BS}

Without the repeating structure,
we may consider the approach in \cite{Wang16},
which is proposed for a narrowband system, with a modification 
for a wideband system (i.e., an OFDM system).

In order to transmit one bit per OFDM symbol (duration) as in \cite{Yang18},
we assume that
\be
\bb_m = \bone s_m,
	\label{EQ:rect}
\ee
where $\bone$ represents a vector of all 1's and
$s_m$ is a binary signal. For example, 
we have $s_m \in \cS = \{-1,+1\}$,
while $\cS =\{0,1\}$ for noncoherent communication in \cite{Wang16}.
Then, the received signal 
(in the frequency-domain) at the receiver becomes
\be
\by_m = (\bH  + \beta \bG s_m) \bx_m + \bn_m.
\ee

To avoid {\bf Difficulty-I} 
(i.e., no carrier estimation is considered at the receiver),
it has to be assumed that $\bH \bx_m$ is unknown. In addition,
$\beta \bG$ is also unknown. In this case, with $s_m \in \{0,1\}$,
the energy detector can be employed with test statistic $||\by_m||^2$
and a decision threshold\footnote{Finding
an optimal threshold requires a careful performance
analysis to minimize error probabilities 
as shown in \cite{Wang16}.
In addition, the optimal threshold depends on statistical properties
of channels, which may not be available in advance.}
as in \cite{Wang16}.
In particular, with $\cS = \{0,1\}$, we have
$$
||\by_m||^2 
=
\left\{
\begin{array}{ll}
||\bH \bx_m + \bn ||^2, & \mbox{if $s_m = 0$} \cr
||(\bH + \beta \bG) \bx_m + \bn ||^2, & \mbox{if $s_m = 1$}.\cr
\end{array}
\right.
$$
Then, the hypothesis test in the energy detector 
can be carried out as follows:
$$
||\by_m||^2 \defh \tau,
$$
where $\tau$ is a decision threshold, and
$\cH_0$ and $\cH_1$ denote the hypotheses for
$s_m = 0$ and $1$, respectively.
Although this approach does not require the carrier estimation and 
the estimation of $\bG$, its performance is usually poor. 
For a better performance, 
we can exploit the knowledge of $\bH$ (as mentioned earlier, 
the receiver is able to estimate $\bH$) to partially
overcome {\bf Difficulty-I} 
and consider
joint estimation and detection.
This approach will be explained in 
Section~\ref{S:MAT}.

\section{Matched-Filter based Transmission
and Low-Complexity Detection}	\label{S:MAT}

In this section, 
in order to overcome the two main difficulties (i.e.,
{\bf Difficulty-I} and {\bf Difficulty-II}),
we discuss a new approach
for modulation in the air and derive a low-complexity joint estimation
and detection method for the receiver.

\subsection{BD's Transmitter Filter with CSI}

Due to the pilot signal transmitted from the legacy OFDM transmitter,
the BD is also able 
to estimate its CSI, $\bG$, or 
the CIR, $\{g_p\}$, in the time-domain. 
For convenience, the CIR is also denoted by $g(t)$. 
Thus, $\{g_p\}$ can be seen as the sampled
(or discrete-time) version of $g(t)$.
Then, the BD can take advantage of known CSI to help the 
receiver in order to improve the performance of
joint estimation and detection.
In particular, we propose the matched-filter based transmission
enforcing the forwarded OFDM signal
by the BD to be coherently received
at the receiver. 

Suppose that 
the time-reversal and complex conjugate
CIR is used as the filter coefficients of the transmitter filter
at the BD, which becomes the \emph{matched-filtering} (to its CSI, $\bG$
or $g(t)$). 
Consequently,
when the BD is to transmit one bit per OFDM symbol,
the baseband signal in the frequency-domain
is represented by
\be
\bb_m = 
\left\{
\begin{array}{rl}
\kappa \bg^*,& \mbox{if Bit $0$ is transmitted} \cr
-\kappa \bg^*,& \mbox{if Bit $1$ is transmitted,} \cr
\end{array}
\right.
	\label{EQ:mfil}
\ee
where $\bg  = [G_0 \ \ldots \ G_{L-1}]^\rT$ and
$\kappa = \frac{\sqrt{L}}{||\bg||}$ so that
$||\bb_m||^2 = L$.
Then, it can be shown that
\begin{align}
\ba_m & = \beta \kappa \bG \bX_m \bg^* s_m \cr
& = \tilde \beta \bV \bx_m s_m,
\ s_m \in \{-1,+1\}. \ 
\end{align}
where $\bV = {\rm diag} (|G_0|^2, \ldots, |G_{L-1}|^2)$
and $\tilde \beta = \beta \kappa$.
From \eqref{EQ:bym},
$\by_m$ becomes
\be
\by_m = (\bH + \tilde \beta \bV s_m) \bx_m + \bn_m.
	\label{EQ:by_ca}
\ee
The  resulting approach with \eqref{EQ:mfil}
is referred to as the channel-aware
matched-filter (CAMF) scheme. On the other hand, for convenience, 
the approach that does not exploit the CSI in \eqref{EQ:rect}
with $s_m \in \{-1,+1\}$
is referred to as the channel-unaware impulse-filter (CUIF) scheme.

\begin{figure}[thb]
\begin{center}
\resizebox{\figwidth}{!}{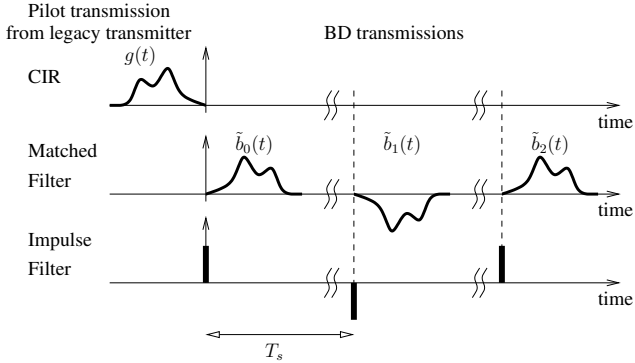}
\end{center}
\caption{Illustrations of the backscattered signals in the time-domain
with two different transmitter filters.}
        \label{Fig:filters}
\end{figure}

In Fig.~\ref{Fig:filters},
the backscattered signals in the time-domain,
which are denoted by $\tilde b_m (t)$ that
can be obtained by the inverse Fourier transform of
$\bb_m$,
are illustrated with two different transmitter filters.
For the CAMF scheme, the BD estimates the CIR, $\{g_p\}$ or $g(t)$.
Then, the impulse response of the transmitter filter
is $g^*(T_s - t)$ and $\tilde b_m (t) = g^*(T_s - t) s_m$,
where $T_s$ denotes the OFDM symbol duration.
On the other hand, in the CUIF scheme,
we have $\tilde b_m (t) = s_m \delta (t)$, i.e., the transmitter
filter is the impulse filter. Thus, the corresponding signal in
the frequency-domain can be represented as in \eqref{EQ:rect}.
Compared with CUIF, CAMF requires more signal processing
operations for the channel estimation and
transmitter filter at the BD due to the matched-filtering.
However, since all the elements of $\bV$, $\{|G_l|^2\}$,
are non-negative real variables, this property can be exploited
at the receiver to overcome {\bf Difficulty-II}.
From this, CAMF can outperform CUIF 
thanks to the matched-filtering.
Note that the signal energy 
(per bit) for backscatter communication, $||\bb_m||^2$,
in the CAMF is the same as that of CUIF scheme, which is $L$.

\subsection{Joint Demodulation and EM Algorithm}

In this subsection, we consider a joint demodulation method 
to estimate $\bV$ (or $\bG$) as well as to detect
backscattered signals
using the EM algorithm \cite{Dempster77} \cite{EM_Book} \cite{ChoiJBook}.

Consider the CAMF scheme. For given $\by_m$, the likelihood
function from \eqref{EQ:by_ca} can be given by
$$
f(\by_m \,|\, \bx_m, s_m, \tilde \beta, \bV)
= C e^{
- \frac{1}{N_0} ||\by_m - (\bH + \tilde \beta \bV s_m) \bx_m||^2 }.
$$
Thus, for given $\{\by_m\}$,
the joint ML problem
to estimate $\{\bx_m, s_m\}$  as well as 
$\{\bV, \tilde \beta\}$ can be formulated as
\be
\max_{\bV, \tilde \beta} \max_{\bx_m \in \cX^L, s_m \in \{-1,+1\}} 
\prod_{m=0}^{M-1} f(\by_m \,|\, \bx_m, s_m, \tilde \beta, \bV).
	\label{EQ:ML0}
\ee
Since the receiver mainly wants to 
detect $s_m$, we can also 
consider the following problem:
\be
\max_{\bV, \tilde \beta} \max_{s_m \in \{-1,+1\}} 
\prod_{m=0}^{M-1} f(\by_m \,|\, s_m, \tilde \beta, \bV),
	\label{EQ:ML1}
\ee
where
\be 
f(\by_m \,|\, s_m, \tilde \beta, \bV) = \sum_{\bx_m \in \cX^L}
f(\by_m \,|\, \bx_m, s_m, \tilde \beta, \bV) \Pr(\bx_m).
\ee
Here, $\Pr(\bx_m)$ is the probability that $\bx_m$ is transmitted
from the legacy OFDM transmitter.
For equally likely OFDM symbols, $\Pr(\bx_m)$ might be a constant
for all $\bx_m \in \cX^L$.
The computational complexity
to solve \eqref{EQ:ML0} 
or \eqref{EQ:ML1} is prohibitively high for large $M$ and $L$.
Thus, we need to resort to low-complexity approaches.

The EM algorithm can be used to iteratively solve
\eqref{EQ:ML1} with low-complexity. 
Suppose that the following ML estimation is to be solved:
\be
\max_{\tilde \beta, \bV} \prod_{m=0}^{M-1} f(\by_m \,|\, \tilde \beta, \bV),
\ee
where 
\be
f(\by_m \,|\,\tilde \beta, \bV) = \sum_{s_m \in \{-1,+1\}}
f(\by_m \,|\, s_m, \tilde \beta, \bV).
\ee
Taking the $s_m$'s as missing variables and
denoting by $\tilde \bV^{(i)}$ the $i$th estimate of 
$\tilde \bV = \tilde\beta \bV$,
the E-step is given by
\be
Q(\tilde \bV\,|\, \tilde \bV^{(i)}) = \sum_m
\uE[ \ln f(\by_m, s_m \,|\, \tilde \bV) 
\,|\, \by_m, \tilde \bV^{(i)}].
	\label{EQ:ES}
\ee
Then, the M-step is given by
\begin{eqnarray}
& \{\tilde \beta^{(i+1)}, \bV^{(i+1)} \} = \argmax_{\tilde \beta, \bV} 
Q(\tilde \bV\,|\, \tilde \bV^{(i)}) \cr
& \mbox{subject to}\ \tilde \beta \in \uC, \ \bV \in \cD^+, &
	\label{EQ:MS}
\end{eqnarray}
where $\cD^+$ represents the set of diagonal matrices whose
elements are nonnegative as $\bV = {\rm diag}(|G_0|^2,
\ldots, |G_{L-1}|^2)$.
Clearly, $\tilde \bV^{(i+1)} = \tilde \beta^{(i+1)} \bV^{(i+1)}$.
It is noteworthy that $\tilde \bV$ has the following constraint:
\begin{align}
\tilde \bV \in
& \tilde \cV = \bigl\{
{\rm diag}(\tilde v_0, \ldots, \tilde v_{L-1})\,|\,
v_l = \tilde \beta a_l, \cr
&  \tilde \beta \in \uC, a_l \ge 0, l = 0,\ldots, L-1 \bigl\}.
	\label{EQ:tVc}
\end{align}
Thus, for any $\tilde \bV = {\rm diag}(\tilde v_0,
\ldots, \tilde v_{L-1}) \in \tilde \cV$,
the phases of the diagonal elements are the same, i.e.,
$$
\angle \tilde v_{l-1} = \angle \tilde v_{l}, \ l = 1, \ldots, L-1, 
$$
which is due to the matched-filter based transmission
in the CAMF scheme.
This constraint on $\tilde \bV$ 
helps find a good estimate
of $\tilde \bV$.

The EM algorithm can also be applied to perform
joint estimation and detection for
the CUIF scheme,
where $\beta \bG$ is to be estimated instead of $\tilde \bV$.
Since the phases of $\beta \bG$ are all different,
there is no constraint such as \eqref{EQ:tVc} to be exploited
for a better estimate of $\tilde \bG$ to overcome {\bf Difficulty-II}.
As a result, as will be shown in Section~\ref{S:Sim},
CUIF cannot perform better than CAMF.

\subsection{Approximations for Low-Complexity Implementations}

In this subsection, 
we discuss approximations for a low-complexity
implementation of the EM algorithm for CAMF.

Let us consider the E-step in \eqref{EQ:ES}.
It can be shown that
\begin{align}
\ln f(\by_m, s_m \,|\, \tilde \bV)  
&= \ln 
\left(f(\by_m \,|\, s_m, \tilde \bV)  \Pr(s_m \,|\, \tilde \bV) \right)\cr
& = 
\ln f(\by_m \,|\, s_m, \tilde \bV) + \ln \Pr(s_m).
\end{align}
Since $s_m$ is equally likely, $\ln  \Pr(s_m)$ becomes a constant.
Thus, it can be shown that
\begin{align}
& \uE[ \ln f(\by_m, s_m \,|\, \tilde \bV) \,|\, \by_m, \tilde \bV^{(i)}]\cr
& = 
\ln f(\by_m \,|\, s_m, \tilde \bV)  \Pr(s_m\,|\, \by_m, \tilde \bV^{(i)})
+ C,
	\label{EQ:Elnf}
\end{align}
where $C$ represents a constant.
Since
$$
||\by_m - (\bH + \tilde \beta \bV s_m) \bx_m||^2
= \sum_{l=0}^{L-1} |y_{l,m} - (H_l + \tilde \beta V_l s_m) x_{l,m}|^2,
$$
we can show that
\begin{align}
\ell_m (s_m) & = \ln f(\by_m \,|\, s_m, \tilde \bV)  \cr
& = \ln \sum_{\bx_m \in \cX} f(\by_m \,|\, \bx_m, s_m, \tilde \bV) + C \cr
& \approx
 \ln  f(\by_m \,|\, \hat \bx_m(s_m), s_m, \tilde \bV) + C \cr
& = 
- \frac{1}{N_0} \sum_{l=0}^{L-1} 
|y_{l,m} - (H_l + \tilde V_l s_m) \hat x_{l,m} (s_m)|^2,
	\label{EQ:Eapp}
\end{align}
where 
\be 
\hat x_{l,m} (s_m)
= \argmin_{x_{l,m} \in \cX} 
|y_{l,m} - (H_l + \tilde V_l^{(i)} s_m) x_{l,m} |^2 .
	\label{EQ:hx}
\ee
In \eqref{EQ:Eapp}, we can see that the estimation of $\bx_m$
is implicitly carried out for the carrier estimation. In particular,
we can see that since $\bH$ is known at the receiver, 
with the estimate of $\bx_m$, the carrier estimation can be performed
to address {\bf Difficulty-I} in joint estimation and detection
using the EM algorithm.

For convenience, let 
$p^{(i)} (s_m) = \Pr(s_m\,|\, \by_m, \tilde \bV^{(i)})
= \frac{e^{\ell_m (s_m)}}{e^{\ell_m (-1)}+ e^{\ell_m (1)} }$.
Then, substituting \eqref{EQ:Eapp} into \eqref{EQ:Elnf} and
\eqref{EQ:ES}, we can show that
\begin{align}
Q(\tilde \bV\,|\, \tilde \bV^{(i)}) \approx
-\frac{1}{N_0} 
 \sum_{l,m} \sum_{s_m \in \{-1,+1\}} \psi_l(s_m) p^{(i)} (s_m) + C,
	\label{EQ:ESA}
\end{align}
where 
\be
\psi_l(s_m) = 
|y_{l,m} - (H_l + \tilde V_l s_m) \hat x_{l,m} (s_m) |^2.
	\label{EQ:psi}
\ee
The complexity to find the approximation of 
$Q(\tilde \bV\,|\, \tilde \bV^{(i)})$ in \eqref{EQ:ESA}
is mainly dependent on the detection of $x_{l,m}$ in 
\eqref{EQ:hx}. For given $\tilde V_l$ and $s_m$,
the complexity to find $\hat x_{l,m} (s_m)$ is proportional
to the size of $\cX$, i.e., $|\cX|$.
Consequently, the resulting complexity 
becomes $O(2 M L |\cX|)$.

The M-step in \eqref{EQ:MS} is to find $\tilde \beta$ and $\bV$, 
which is not easy to solve due to the constraints on $\tilde \beta$
and $\bV$. However,
if 
$\tilde \bV = \tilde \beta \bV$ is to be found (rather than $\tilde \beta$ and
$\bV$) without the constraint in \eqref{EQ:tVc}, 
\eqref{EQ:MS} is reduced to a minimization 
of a quadratic function of $\tilde V_l$.
In particular, from \eqref{EQ:ESA}, it can be shown that
$$
Q(\tilde \bV\,|\, \tilde \bV^{(i)}) \approx
\sum_{l=0}^{L-1}
Q_l (\tilde V_l\,|\, \tilde \bV^{(i)})+ C,
$$
where
\be
Q_l (\tilde V_l\,|\, \tilde \bV^{(i)})= 
-\frac{1}{N_0} 
 \sum_{m} \sum_{s_m \in \{-1,+1\}} \psi_l(s_m) p^{(i)} (s_m).
\ee
Since $\psi_l (s_m)$ is a quadratic function of $\tilde V_l$
as shown in \eqref{EQ:psi}, 
$Q_l (\tilde V_l\,|\, \tilde \bV^{(i)})$ is also a quadratic function
of $\tilde V_l$.
Therefore, $\tilde V_l$ that
maximizes $Q_l (\tilde V_l\,|, \tilde \bV^{(i)})$ can be readily
obtained, which is denoted by $\tilde \nu_l^{(i+1)}$.
To impose the constraint in \eqref{EQ:tVc},
from $\{\tilde \nu_{l}^{(i+1)}\}$,
$\bV^{(i+1)}$ can be found as
\be
\tilde 
\bV^{(i+1)} = e^{j \hat \theta} {\rm diag}(
|\tilde v_0^{(i+1)}|, \ldots,
|\tilde v_{L-1}^{(i+1)}|) \in \tilde \cV,
	\label{EQ:Vi1}
\ee
where $\theta$ is the average angle that is obtained
as
$$
\hat \theta = \angle \left(\frac{1}{L} \sum_l 
\tilde v_l^{(i+1)} \right).
$$

In summary, the joint estimation and detection can be
performed as follows.
\begin{itemize}
\item[{\bf C0)}] 
{\bf Input}: $\{\by_m\}$ and $\bH$.
Set an initial value for
$\tilde \bV^{(i)}$ with $i = 0$,
\item[{\bf C1)}] Compute $p^{(i)} (s_m) = \Pr(s_m\,|\, \by_m,
\tilde \bV^{(i)})$ for all $m$ using \eqref{EQ:Eapp}.
\item[{\bf C2)}] Find $Q_l (\tilde V_l\,|\, \tilde \bV^{(i)})$ for all $l$
and update $\bV$ as in \eqref{EQ:Vi1}.
\item[{\bf C3)}] If $||\tilde \bV^{(i+1)} - \tilde \bV^{(i)} ||_{\rm F}^2
< \epsilon$ or $i > N_{\rm iter}$ stop. Otherwise,
move to {\bf C1)} with $i \Rightarrow  i+1$.
\end{itemize}
Here, $\epsilon > 0$ is a pre-determined threshold
and $N_{\rm iter}$ represents the maximum number of iterations.

Note that 
the complexity for the M-step is also linearly proportional 
to $L$ (as $Q_l (\tilde V_l|\tilde \bV^{(i)})$ is a quadratic
function of $\tilde V_l$ as mentioned earlier). 
Thus, the complexity per iteration 
(for both E- and M-steps) remains $O(ML |\cX|)$.
Consequently, the overall complexity of the EM
algorithm for joint estimation and detection becomes
$O(N_{\rm iter} ML|\cX|)$. 
As will be shown in Section~\ref{S:Sim}, 
$N_{\rm iter}$
is small (e.g., 5 iterations are required in general). 
Clearly, we can claim that joint estimation and detection
can be carried out with a complexity 
(linearly) proportional to $ML$.

\section{Performance Analysis}	\label{S:PA}

In this section,
we analyze the performance of the CAMF and CUIF schemes
under certain ideal conditions. 
In particular, throughout this section,
we assume that the receiver not only
knows its CSI, $\bH$, but also BD's CSI, i.e., $\bG$.
In practice, since $\bG$ is to be estimated
in joint estimation and detection, the assumption
leads to optimistic results.

\subsection{With Known OFDM Symbols}

For tractable analysis,
we consider the following assumption.
\begin{itemize}
\item[{\bf A0})] The receiver knows the OFDM symbols, i.e.,
$$
\hat x_{l,m} (s_m) = x_{l,m}.
$$
\end{itemize}
That is, the OFDM symbol is correctly known at the receiver, which
results in an optimistic performance for the detection 
of the backscattered signal.
Denote by $\uP_{\rm cuif}$ 
the probability of bit error or bit error rate (BER)
of CUIF.
Then, we have
\begin{align}
\uP_{\rm cuif} 
& = \Pr( ||\by - (\bH + \beta \bG) \bx ||^2
\ge ||\by - (\bH - \beta \bG) \bx ||^2) \cr
& = \uE [ \uP_{\rm cuif} (\bG)],
	\label{EQ:Pif1}
\end{align}
where
\be
\uP_{\rm cuif} (\bG)
=\cQ \left( \sqrt\frac{||2 \beta \bG \bx||^2}{2 N_0} \right).
\ee
To derive a closed-form expression for the BER, 
we consider the following assumptions.

\begin{itemize}
\item[{\bf A1)}] 
The coefficients of the CIR from the legacy OFDM 
transmitter to the BD are independent and
\be
g_p \sim \cC \cN (0, \sigma_{g,p}^2),
\ee
i.e., multipath Rayleigh fading is considered for the
channel from the legacy OFDM transmitter to the BD.
\item[{\bf A2)}] The amplitude of
$x_{l,m}$ is constant, i.e., $|x_{l,m}|$ is the same
for all $l$ and $m$
and $|x_{l,m}|^2 = E_x$. For example,
4-quadrature amplitude modulation (QAM) can be considered.
\end{itemize}

Under the assumptions of {\bf A0}, {\bf A1}, and {\bf A2},
the conditional BER of CUIF can be given by
\begin{align}
\uP_{\rm cuif} (\bG)
& = \Pr ( ||\bn ||^2 \ge ||2 \beta \bG \bx + \bn||^2\,|\, \bG) \cr
& = \cQ
\left(
\sqrt\frac{2 |\beta|^2 \sum_l |G_l x_{l}|^2 }{N_0} 
\right).
	\label{EQ:cb1}
\end{align}
Since
$\sum_l  |G_l x_{l}|^2 = \sum_l  |G_l|^2 E_x 
= L E_x \sum_{p =0}^{P_g-1} |g_p|^2$,
\eqref{EQ:cb1} is rewritten as
\be
\uP_{\rm cuif} (\bG) = 
\cQ \left( \sqrt{2 |\beta|^2 L \gamma_x \sum_p |g_p |^2 } \right).
	\label{EQ:cb2}
\ee
Thus, the BER of CUIF can be found as
\begin{align}
\bar \uP_{\rm cuif} (L) & = \uE \left[
\cQ \left( \sqrt{2 L |\beta|^2 \gamma_x \sum_p |g_p |^2 } \right)
\right] \cr
& \approx 
\frac{1}{12} 
\prod_{p=0}^{P_g-1} \frac{1}{1+L |\beta|^2 \gamma_x \sigma_{g,p}^2} \cr
& \quad + 
\frac{1}{4} 
\prod_{p=0}^{P_g-1} \frac{1}{1+
\frac{4}{3}L |\beta|^2 \gamma_x \sigma_{g,p}^2},
	\label{EQ:Pcuif}
\end{align}
where $\gamma_x = \frac{E_x}{N_0}$.
The approximation in \eqref{EQ:Pcuif} is due to 
\cite{Chiani02}, where $\cQ(x)$ can be approximated as
$\cQ(x) \approx \frac{1}{12} e^{- \frac{x^2}{2}}
+ \frac{1}{4} e^{-  \frac{2 x^2}{3}}$.

If $\sigma_{g,p}^2 = \sigma_g^2$ for all $p$,
from \cite{ProakisBook} \cite{SimonBook00},
the BER of CUIF can also be expressed as
\be
\bar \uP_{\rm cuif} (L)= 
\left( \frac{1-\mu}{2} \right)^{P_g}
\sum_{p=0}^{P_g-1} 
\binom{P_g-1+p}{p}
\left( \frac{1+\mu}{2} \right)^{p},
	\label{EQ:ep}
\ee
where
$\mu = \sqrt{
\frac{L |\beta|^2 \gamma_x \sigma_g^2}
{1+ L |\beta|^2 \gamma_x \sigma_g^2}}$.
From \eqref{EQ:Pcuif} and \eqref{EQ:ep}, it is clear that the diversity 
gain is $P_g$ and the BER decreases with $L$ as the bit
energy (of scattered signals) increases. 
In addition, the BER decreases with the reflection coefficient,
$\alpha$, or $|\beta|^2$.

Similarly, we can have the conditional BER of CAMF
as follows:
\begin{align}
\uP_{\rm camf} (\bG)
=
\cQ \left( \sqrt\frac{||2 \tilde \beta \bV \bx||^2}{2 N_0} \right) ,
	\label{EQ:Pmf1}
\end{align}
while the BER can be obtained by taking
the mean with respect to $\bG$, which is denoted by
$\bar \uP_{\rm camf} (L) =\uE [ \uP_{\rm camf} (\bG)]$.

\begin{mytheorem}	\label{T:ineq}
Under the assumptions of {\bf A0}, {\bf A1}, and {\bf A2},
it can be shown that
\be
\uP_{\rm camf} (\bG) \le \uP_{\rm cuif} (\bG).
	\label{EQ:ineq}
\ee
We have the equality in \eqref{EQ:ineq} if
the $|G_l|^2$'s are the same for all $l$.
\end{mytheorem}
\begin{IEEEproof}
From \eqref{EQ:Pif1} and \eqref{EQ:Pmf1},
for the inequality in \eqref{EQ:ineq},
it is sufficient to show that
\be
||\tilde \beta \bV \bx||^2 \ge ||\beta \bG \bx||^2
= |\beta|^2 E_x L \sum_{p=0}^{P_g - 1} |g_p|^2.
	\label{EQ:ineq1}
\ee
Since $\kappa = \frac{\sqrt{L}}{||\bg||}$,
it can be shown that
\begin{align}
||\tilde \beta \bV \bx||^2 
& =|\beta|^2 |\kappa |^2 E_x \sum_l |G_l|^4 \cr
&= \frac{|\beta|^2 L E_x \sum_l |G_l|^4 }{\sum_l |G_l|^2} =
\frac{|\beta|^2 E_x \sum_l |G_l|^4 }{\sum_p |g_p|^2}.
	\label{EQ:bVi}
\end{align}
Using the Cauchy-Schwarz inequality,
we can have the following inequality for a sequence
$\{z_1, \ldots, z_n\}$:
\be
|\sum_{i=1}^n z_i |^2 \le n\sum_{i=1}^n |z_i|^2.
	\label{EQ:iz}
\ee
This inequality
can be applied to \eqref{EQ:bVi} and leads to
\begin{align}
||\tilde \beta \bV \bx||^2 
& \ge 
|\beta|^2 E_x \frac{| \sum_l |G_l|^2 |^2}{L \sum_p |g_p|^2} \cr
& = 
|\beta|^2 E_x L \sum_p |g_p|^2,
\end{align}
which is identical to \eqref{EQ:ineq1}. 

For the equality in \eqref{EQ:ineq1},
let us consider the inequality in \eqref{EQ:iz},
where $z_i = |G_i|^2$. If all $z_i$'s are the same, 
the inequality in \eqref{EQ:iz} becomes the equality,
which implies that $||\tilde \bV \bx||^2 = ||\beta \bG \bx||^2$
and 
$\uP_{\rm camf} (\bG) = \uP_{\rm cuif} (\bG)$. This completes the proof.
\end{IEEEproof}

Although the BER of CAMF is lower than that of CUIF as in \eqref{EQ:ineq}, 
the gap may not be significant if the $|G_l|^2$'s are not significantly
varying. 
In particular, if $|G_l|^2 \approx A$ for all $l$
(e.g., near flat fading),
as in Theorem~\ref{T:ineq}, the BER of CAMF might be close to
that of CUIF under ideal conditions
(i.e., known $\bx_m$ and $(\bH, \bG)$). 


\subsection{With Unknown OFDM Symbols}

Since the assumption of {\bf A0}
leads to an exceedingly optimistic performance,
we may need to take into account errors in detecting OFDM symbols.
In this subsection, we consider an approach
to find the performance of
CUIF\footnote{Due to \eqref{EQ:ineq},
it is expected that the performance  of CAMF is slightly better than
that of CUIF in the case of unknown OFDM symbols as well.}
with detection errors of OFDM symbols at the receiver.

Since the OFDM symbol detection
and backscattered signal detection
are interrelated, the analysis with unknown OFDM symbols
becomes involved. To avoid this difficulty,
we consider a two-step approach.
In the first step, we find the detection error probability
of OFDM symbols under the assumption that
the backscattered signal becomes Gaussian interference.
In the second step, we assume that the received
signals corresponding to erroneously detected
OFDM symbols are not used for the detection of backscattered 
signals. This implies that the number of subcarriers, $L$,
decreases and results in a low SNR in the
BER expression for CUIF, i.e., \eqref{EQ:Pcuif}.

In the first step, since the backscattered signals
are assumed to be Gaussian interference,
under the assumption of {\bf A1},
the variance of the background noise and interference becomes
\be
K_0 = N_0 + |\beta|^2  E_x \sum_{p =0}^{P_g - 1} \sigma_{g,p}^2.
\ee
Since $x_{l,m}$ is a 4-QAM symbol (according to the assumption
of {\bf A2}),
the BER of the OFDM symbol detection
becomes 
\be
\bar \uP_{\rm ber} = \uE \left[
\cQ \left(\sqrt\frac{2 |H_l|^2}{K_0} \right)
\right].
\ee
As in {\bf A1}, we can also consider
multipath Rayleigh fading for $\bH$, i.e., 
the $h_p$'s are independent and
\be
h_p \sim \cC \cN (0, \sigma_{h,p}^2).
\ee
Then, since $|H_l|^2$ becomes an exponential random variable,
according to \cite{ProakisBook},
the (average) BER of each OFDM data symbol, $x_{l,m}$, is given by
\be
\bar \uP_{\rm ber} = \frac{1}{2} \left(
1 - \sqrt{\frac{\gamma_h}{1+ \gamma_h}}
\right),
\ee
where $\gamma_h = \frac{\sum_{p=0}^{P_h - 1} \sigma_{h,p}^2}{K_0}$.
If each bit error in $x_{l,m}$ is assumed to be independent,
we have the symbol error rate (SER) of $x_{l,m}$ as follows
$$
\bar \uP_{\rm ser} = 1 - (1 - \bar \uP_{\rm ber})^2.
$$
This is the probability that the received signal through
a subcarrier is not taken into account due to erroneous
detection of $x_{l,m}$ for the
backscattered signal detection.
Thus, the probability that the number of subcarriers
that can have correct OFDM symbol detection 
is $k$ given by
\be
\Pr(\hat L = k) = \binom{L}{k} (1 - \bar \uP_{\rm ser})^k 
 \uP_{\rm ser}^{L-k },
\ee
and the BER of CUIF with unknown OFDM symbols
can be given by
\be
\bar \uP_{\rm cuif}  = \sum_{k=0}^L 
\Pr(\hat L = k) \bar \uP_{\rm cuif} (k).
	\label{EQ:epX}
\ee
Note that since $\beta \bG$ is assumed
to be known at the receiver, \eqref{EQ:epX} 
can also be seen as a lower-bound on the actual BER for the detection
of backscattered signals by joint estimation and detection
where $\beta \bG$ is to be estimated.
In addition, \eqref{EQ:epX} may also be used for CAMF
as an ideal performance, because 
the difference between $\uP_{\rm camf} (\bG)$ 
and $\uP_{\rm cuif} (\bG)$ may not be too large as mentioned earlier.

\section{Simulation Results}	\label{S:Sim}

In this section, we present simulation results with
Rayleigh multipath channels for $\{h_p\}$ and $\{g_p\}$,
where each channel coefficient is an independent zero-mean
CSCG random variable and
$\sigma_{h,p}^2 = \sigma_{g,p}^2 = \frac{1}{P}$, $p = 0,\ldots, P-1$.
Here, $P_h = P_g = P$. 
The channel coefficient from the BD to the receiver is 
assumed to be $f = e^{j \theta}$ for all the subcarriers,
where $\theta$ is a random phase.
In addition, we consider 4-QAM with $\cX = \{\pm 1 \pm j \}$.
Since $E_x = 2$, the SNR (for OFDM data symbols, $x_{l,m}$)
or $E_{\rm b}/N_0$
is given as $\frac{1}{N_0}$, where $E_{\rm b}$ is the bit energy 
(which is a half of $E_x$ for 4-QAM).

For the EM algorithm, we consider a fixed number of iterations.
For the initial values of the EM algorithm,
we assume $p^{(0)} (s_m = +1) =p^{(0)} (s_m = -1) = \frac{1}{2}$
for all $m$ 
and $\tilde V_l^{(0)} = \alpha e^{j \hat \theta^{(0)}}$ for all $l$,
where $\hat \theta^{(0)}$ is an initial estimate of $\theta = \angle (f)$.
Note that the convergence of the EM algorithm 
depends on the initial values \cite{EM_Book}.
Thus, some initial values should be carefully chosen,
in particular, the initial values of
$\tilde V_l$, i.e., $\tilde V_l^{(0)}$, for CAMF.
In this section, we assume that
$$
|\hat \theta^{(0)} - \theta| \le \frac{\pi}{4},
$$
which means that the receiver is able to determine the phase of
the channel coefficient, $f$, up to $\pm \frac{\pi}{4}$ difference through
a rough initial estimate (to this end, the first bit of BD might be
a pilot bit that is known to the receiver).

\begin{figure}[thb]
\begin{center}
\includegraphics[width=\figwidth]{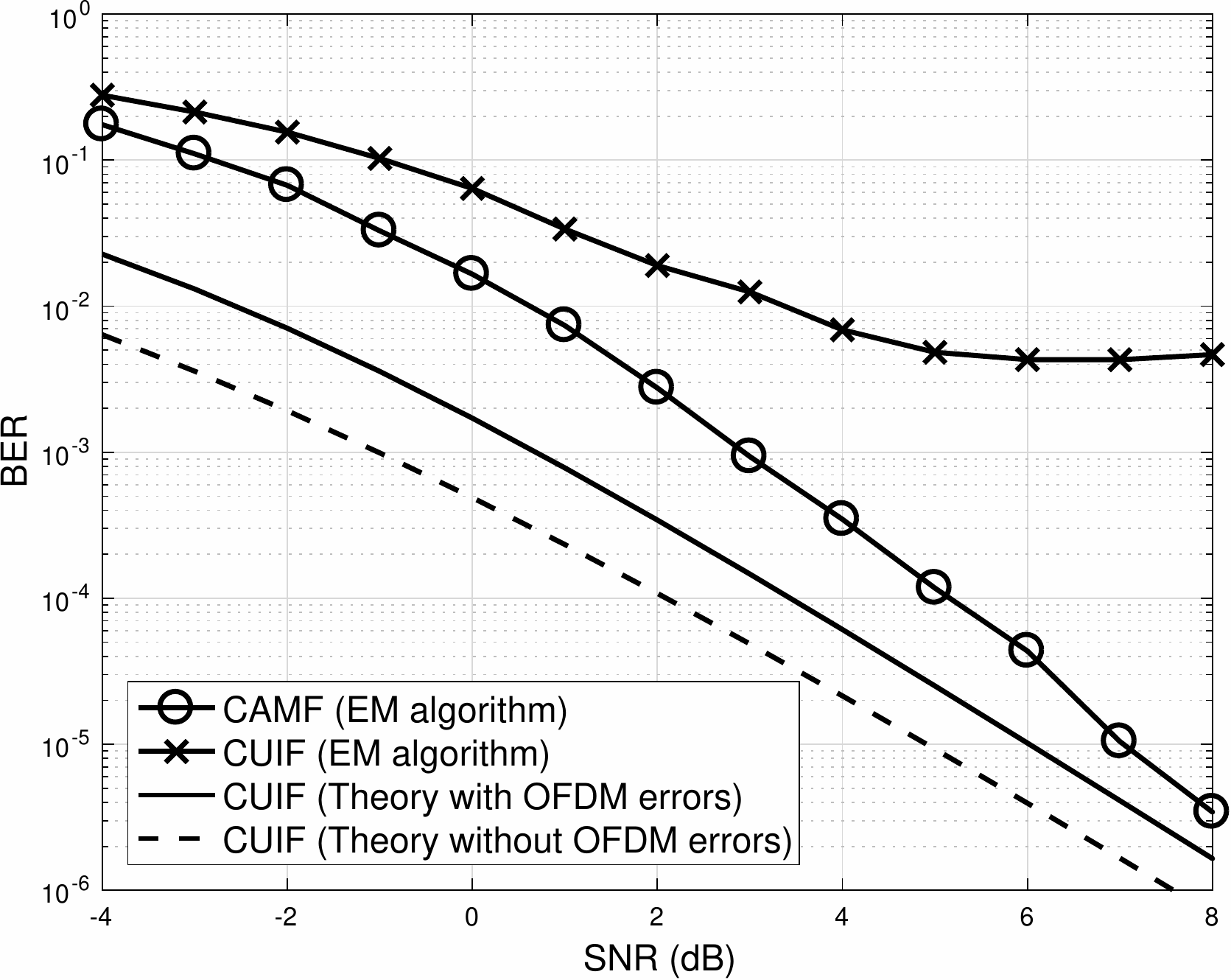} 
\end{center}
\caption{BERs of CAMF and CUIF as functions of SNR with
$L = 32$, $P = 4$, $M = 100$, and $\alpha^2 = 0.2$. 
For the EM algorithm, we have $N_{\rm iter} = 5$.}
        \label{Fig:plt1}
\end{figure}

Fig.~\ref{Fig:plt1}
shows the BERs of CAMF and CUIF 
as functions of SNR with
$L = 32$, $P = 4$, $M = 100$, and $\alpha^2 = 0.2$. 
For the EM algorithm, $N_{\rm iter}$ is set to 5.
The theoretical BERs
of CUIF without and with OFDM symbol errors in
\eqref{EQ:ep} and 
\eqref{EQ:epX}, respectively, are also shown.
Since they are obtained under the assumption that $\bG$ is perfectly
known at the receiver, both might be seen as lower-bounds.
The BERs of CAMF and CUIF obtained from the EM algorithm 
decrease with the SNR, 
while CAMF performs better than CUIF as expected.
It is noteworthy that
the BER curve of CUIF has an error floor 
at a high SNR (i.e., $\ge 5$dB), which might be due to the estimation
error of $\bG$. 
Since the BER of CAMF from the EM algorithm 
becomes close to the theoretical BER of CUIF
with OFDM symbol errors in \eqref{EQ:epX} as the SNR increases,
we can confirm that the estimation of $\tilde \bV$ can be 
reliably carried out in CAMF thanks to the matched-filtering
at BD.

\begin{figure}[thb]
\begin{center}
\includegraphics[width=\figwidth]{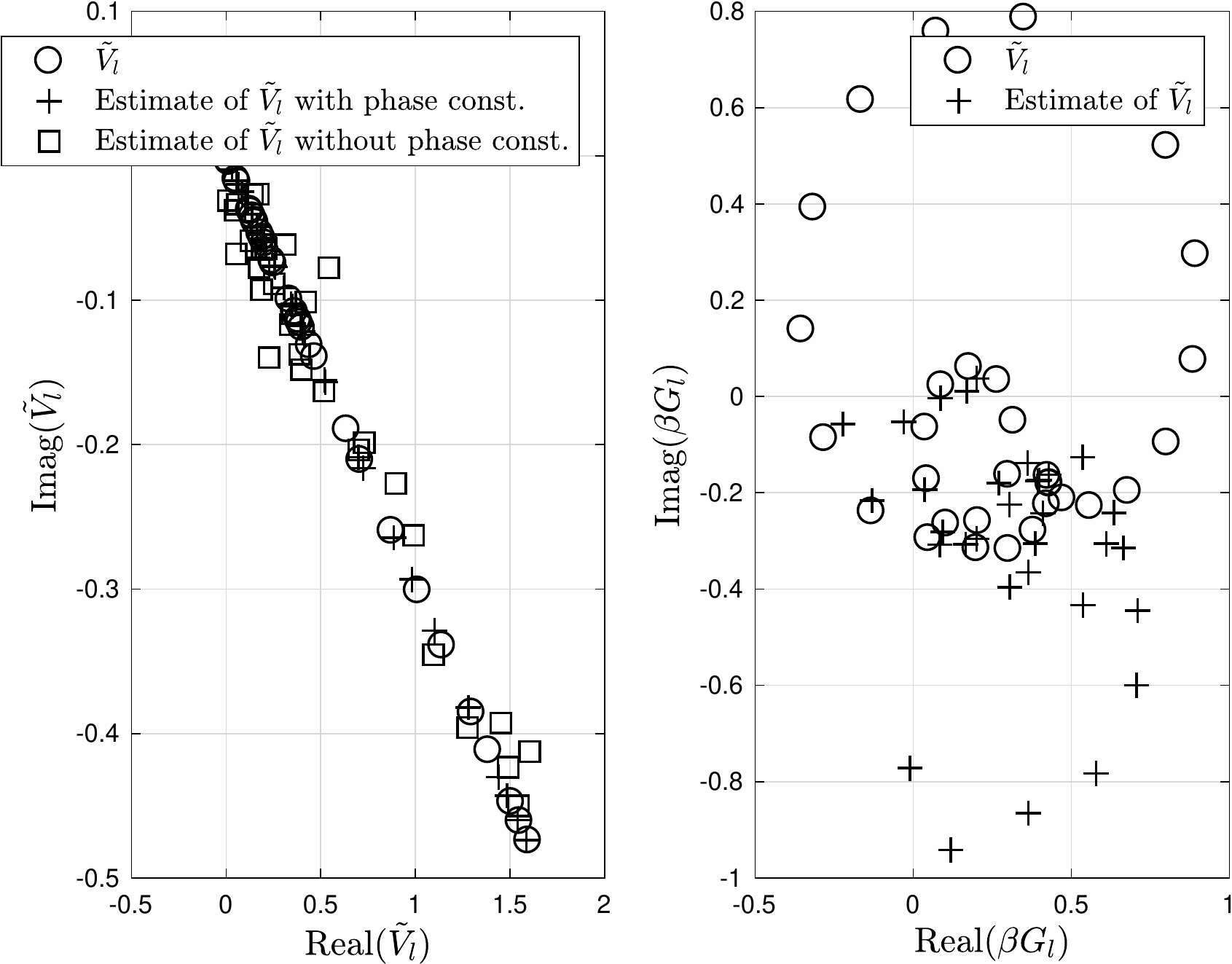} \\
\hskip 0.5cm (a) \hskip 3.5cm (b) \\
\end{center}
\caption{The estimates of $\tilde V_l$ and
$\beta G_l$ for CAMF and CUIF, respectively,
from the EM algorithm with $N_{\rm iter} = 5$
when $L = 32$, $P = 4$, $M = 100$, 
SNR $= 4$dB, and $\alpha^2 = 0.2$.}
        \label{Fig:plt_ph}
\end{figure}

As mentioned earlier, CAMF can provide a better performance
than CUIF because the estimation of $\tilde \bV$ in CAMF
can be more reliable than that of $\beta \bG$ in CUIF.
To see this, in Fig.~\ref{Fig:plt_ph}, we show
the estimates of $\tilde \bV$
and $\beta \bG$ for CAMF and CUIF,
respectively,
from the EM algorithm with $N_{\rm iter} = 5$
when $L = 32$, $P = 4$, $M = 100$, 
SNR $= 4$dB, and $\alpha^2 = 0.2$.
Exploiting the constraint that
the phases of $\tilde V_l$'s are the same
(thanks to the matched-filtering at BD),
it is shown that the EM algorithm for CAMF
can provide a more reliable estimate of $\tilde \bV$
than that of $\beta \bG$ for CUIF,
which results in a performance difference in Fig.~\ref{Fig:plt1}.

\begin{figure}[thb]
\begin{center}
\includegraphics[width=\figwidth]{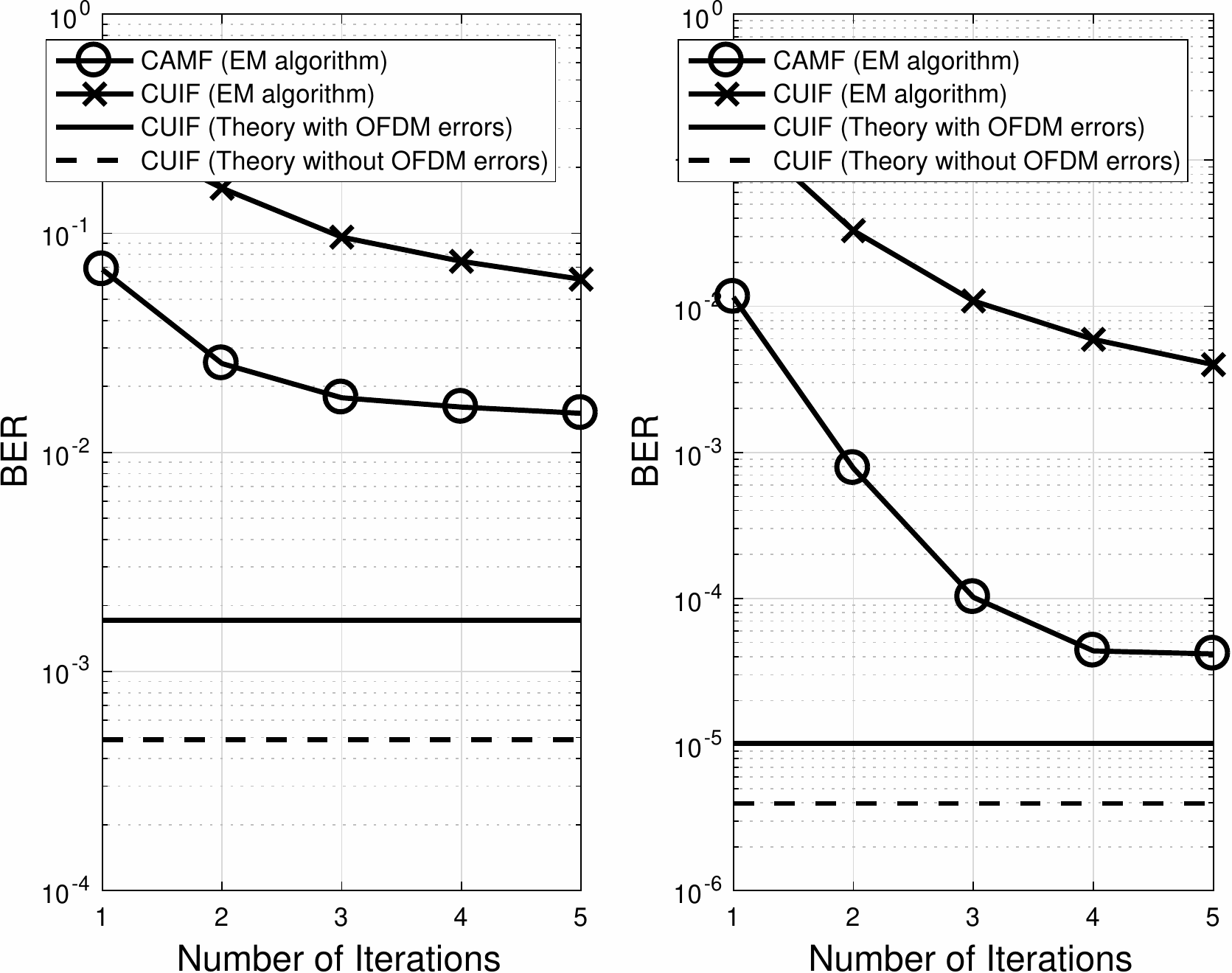} \\
\hskip 0.5cm (a) \hskip 3.5cm (b) \\
\end{center}
\caption{BERs of CAMF and CUIF as functions of the number
of iterations in the EM algorithm with
$L = 32$, $P = 4$, $M = 100$, and $\alpha^2 = 0.2$: 
(a) SNR $= 0$dB; (b) SNR $= 6$dB.}
        \label{Fig:plt2}
\end{figure}

In order to see the performance improvement by increasing the number
of iterations, $N_{\rm iter}$, in the EM algorithm, 
the BERs of CAMF and CUIF are obtained as functions
of $N_{\rm iter}$ and shown in Fig.~\ref{Fig:plt2}
when $L = 32$, $P = 4$, $M = 100$, and $\alpha^2 = 0.2$.
It is shown that $N_{\rm iter} = 5$ might be sufficient for CAMF
(even at a high SNR, e.g., 6dB, as shown in
Fig.~\ref{Fig:plt2} (b)).

\begin{figure}[thb]
\begin{center}
\includegraphics[width=\figwidth]{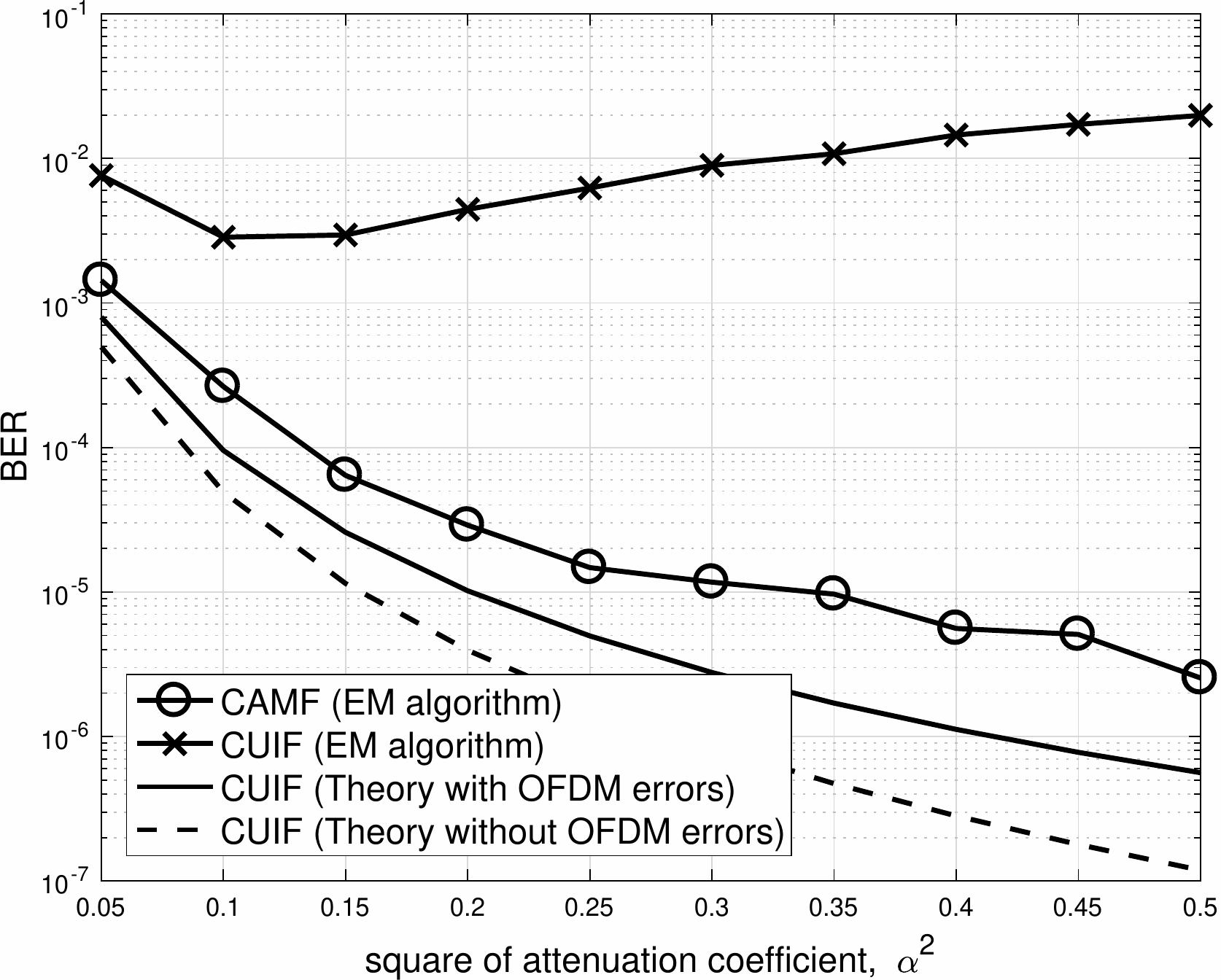} 
\end{center}
\caption{BERs of CAMF and CUIF as functions of $\alpha^2$ with
$L = 32$, $P = 4$, $M = 100$, and SNR $= 6$dB. 
For the EM algorithm, we have $N_{\rm iter} = 5$.}
        \label{Fig:plt4}
\end{figure}

Fig.~\ref{Fig:plt4} shows the BERs
of CAMF and CUIF as functions of $\alpha^2$ with
$L = 32$, $P = 4$, $M = 100$, and SNR $= 6$dB. 
For the EM algorithm, we have $N_{\rm iter} = 5$.
In general, since the signal strength of
scattered signals increases with $\alpha^2$,
we expect a better performance for a larger $\alpha^2$
provided that the receiver is able to perform
a reliable carrier estimation.
This can happen with CAMF 
as the EM algorithm can perform reasonably well for
joint estimation and detection. However, with CUIF,
due to the difficulty in estimating $\beta \bG$,
joint estimation and detection cannot be properly carried out
at the receiver and the performance becomes worse as $\alpha^2$
increases.

\begin{figure}[thb]
\begin{center}
\includegraphics[width=\figwidth]{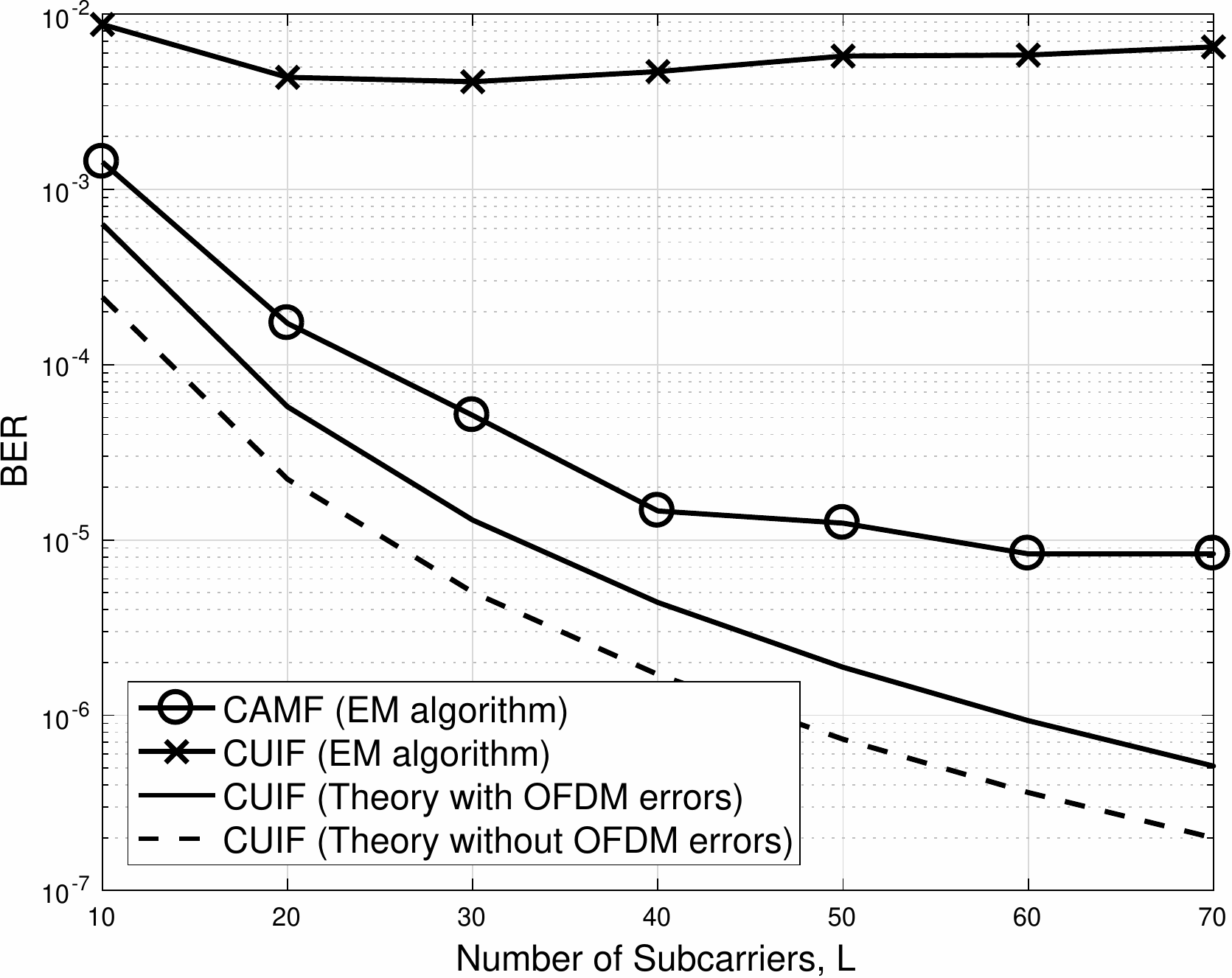} 
\end{center}
\caption{BERs of CAMF and CUIF as functions of 
the length of OFDM symbol, $L$, with
$P = 4$, $\alpha^2 = 0.2$, $M = 100$, and SNR $= 6$dB. 
For the EM algorithm, we have $N_{\rm iter} = 5$.}
        \label{Fig:plt5}
\end{figure}

In Fig.~\ref{Fig:plt5}, we show 
the BERs of CAMF and CUIF as functions of 
the length of OFDM symbol, $L$, with
$P = 4$, $\alpha^2 = 0.2$, $M = 100$, and SNR $= 6$dB. 
For the EM algorithm, $N_{\rm iter}$ is set to 5.
Since $L$ is the bit energy for backscattered signals,
a better performance is expected as $L$ increases.
With CAMF, since joint estimation and detection
can be successfully performed, it is shown that
the BER decreases with $L$. However, with CUIF, 
due to the difficulty to estimate the phases of $\beta \bG$,
the performance improvement by increasing $L$ cannot be seen.

\begin{figure}[thb]
\begin{center}
\includegraphics[width=\figwidth]{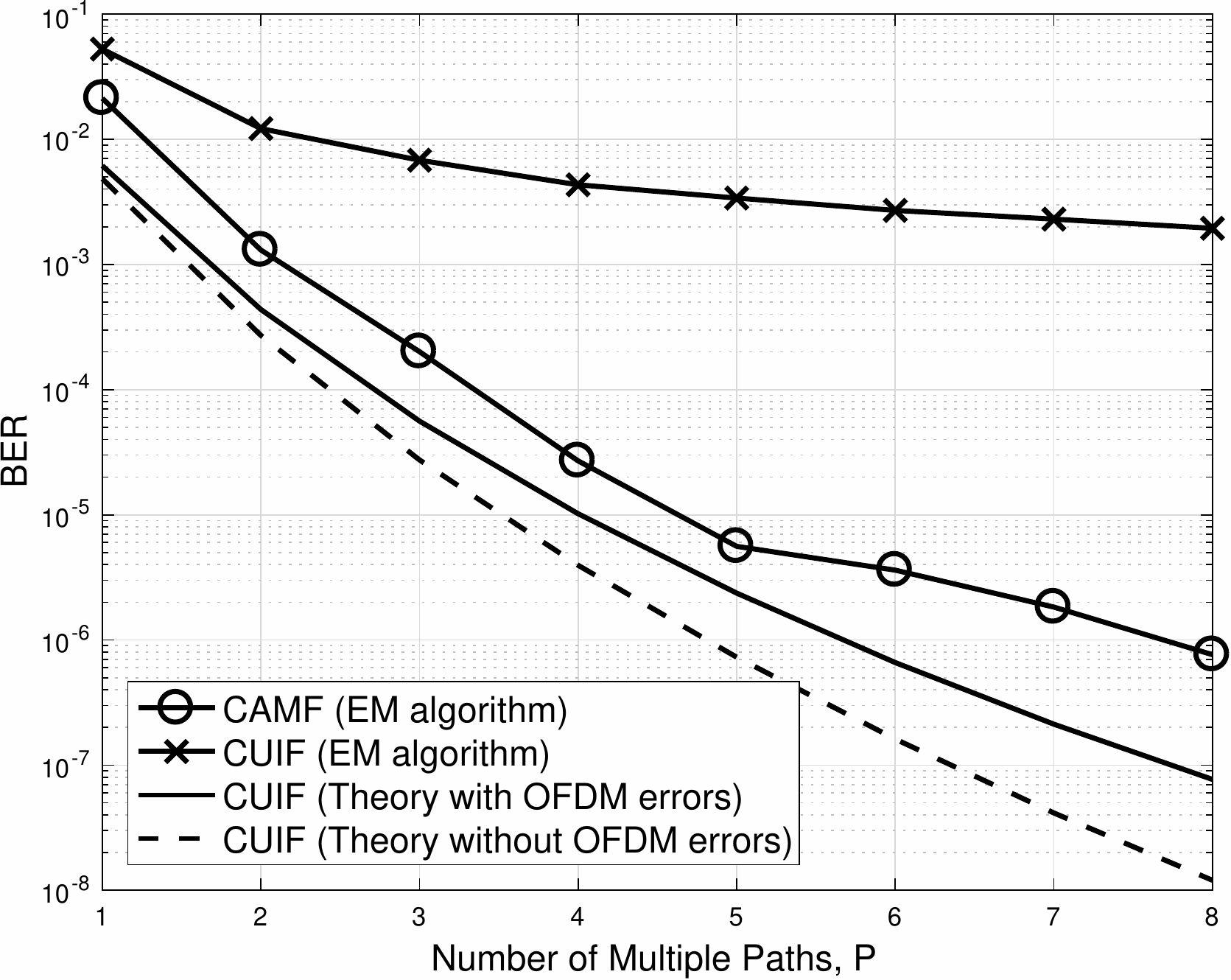} 
\end{center}
\caption{BERs of CAMF and CUIF as functions of 
the length of CIR, $P$, with
$L = 32$, $\alpha^2 = 0.2$, $M = 100$, and SNR $= 6$dB. 
For the EM algorithm, we have $N_{\rm iter} = 5$.}
        \label{Fig:plt6}
\end{figure}

The impact of the number of multiple paths, $P$,
on the BER is shown in Fig.~\ref{Fig:plt6}
with $L = 32$, $\alpha^2 = 0.2$, $M = 100$, and SNR $= 6$dB. 
For the EM algorithm, we have $N_{\rm iter} = 5$.
Since the diversity gain increases with $P$
(which is shown in 
\eqref{EQ:Pcuif} and \eqref{EQ:ep}),
we expect a lower BER as $P$ increases, which is also confirmed
by the simulation results in Fig.~\ref{Fig:plt6}.

As mentioned earlier, in general, 
it is difficult for 
the receiver to estimate BD's CSI, $\beta \bG$,
as it is not directly observable (due to OFDM symbols, $\bx_m$, multiplied
to $\bG$).
As a result,
the performance of CUIF is not close to 
the theoretical one in \eqref{EQ:epX}
(with known BD's CSI) as shown 
in Figs.~\ref{Fig:plt1}-\ref{Fig:plt6}.
On the other hand, in those simulation results,
it is shown that
the theoretical BER of CUIF in \eqref{EQ:epX}
can help predict the performance of the proposed CAMF
scheme, although it serves as a lower-bound. 
This implies that the joint estimation and detection
at the receiver is able to 
reliably estimate BD's CSI (i.e., $\tilde \bV$)
in CAMF thanks to the matched-filtering
(as shown in Fig.~\ref{Fig:plt_ph}).
Therefore, we can expect a reliable detection of backscattered
signals using CAMF at the expense of more signal processing
operations at both the BD and receiver.

\section{Concluding Remarks}	\label{S:Conc}

In this paper, we proposed a novel modulation scheme, CAMF, for
AmBC using OFDM carrier that can effectively exploit BD's CSI
to help overcome the difficulty
in performing joint estimation and detection 
at a receiver.
Using the matched-filtering (to BD's CSI),
it became possible to enforce the forwarded OFDM signal
by the BD to be coherently received
at the receiver. As a result, a reliable estimate of BD's CSI
has been available at the receiver
and the backscattered signals have also been reliably detected.
For low-complexity joint estimation and detection,
the EM algorithm was considered with approximations.
From simulation results, we have demonstrated that
the performance of CAMF can be close to the theoretical one
(which is obtained with known BD's CSI) thanks to the matched-filtering,
while the performance became poor without the matched-filtering.

\bibliographystyle{ieeetr}
\bibliography{backscat}

\end{document}

%% file: filters.pdf_t
\begin{picture}(0,0)%
\includegraphics{filters.pdf}%
\end{picture}%
\setlength{\unitlength}{3947sp}%
\begingroup\makeatletter\ifx\SetFigFont\undefined%
\gdef\SetFigFont#1#2#3#4#5{%
  \reset@font\fontsize{#1}{#2pt}%
  \fontfamily{#3}\fontseries{#4}\fontshape{#5}%
  \selectfont}%
\fi\endgroup%
\begin{picture}(6414,3693)(61,-3589)
\put(1276,-511){\makebox(0,0)[lb]{\smash{{\SetFigFont{12}{14.4}{\rmdefault}{\mddefault}{\updefault}{\color[rgb]{0,0,0}$g(t)$}%
}}}}
\put(2701,-3511){\makebox(0,0)[lb]{\smash{{\SetFigFont{12}{14.4}{\rmdefault}{\mddefault}{\updefault}{\color[rgb]{0,0,0}$T_s$}%
}}}}
\put(2401,-1411){\makebox(0,0)[lb]{\smash{{\SetFigFont{12}{14.4}{\rmdefault}{\mddefault}{\updefault}{\color[rgb]{0,0,0}$\tilde b_0 (t)$}%
}}}}
\put(3901,-1411){\makebox(0,0)[lb]{\smash{{\SetFigFont{12}{14.4}{\rmdefault}{\mddefault}{\updefault}{\color[rgb]{0,0,0}$\tilde b_1 (t)$}%
}}}}
\put(5401,-1411){\makebox(0,0)[lb]{\smash{{\SetFigFont{12}{14.4}{\rmdefault}{\mddefault}{\updefault}{\color[rgb]{0,0,0}$\tilde b_2 (t)$}%
}}}}
\end{picture}%

%% file: final.bbl
\begin{thebibliography}{10}

\bibitem{ITU_IoT}
ITU-T, {\em Y.2060: Overview of the Internet of things}, June 2012.

\bibitem{Keoh14}
S.~L. Keoh, S.~S. Kumar, and H.~Tschofenig, ``Securing the {I}nternet of
  {T}hings: A standardization perspective,'' {\em IEEE Internet of Things
  Journal}, vol.~1, pp.~265--275, June 2014.

\bibitem{Gollakota14}
S.~Gollakota, M.~S. Reynolds, J.~R. Smith, and D.~J. Wetherall, ``The emergence
  of {RF}-powered computing,'' {\em Computer}, vol.~47, pp.~32--39, Jan 2014.

\bibitem{Dobkin12}
D.~M. Dobkin, {\em The RF in RFID, Second Edition: UHF RFID in Practice}.
\newblock Newton, MA, USA: Newnes, 2nd~ed., 2012.

\bibitem{Boyer14}
C.~Boyer and S.~Roy, ``Backscatter communication and {RFID}: Coding, energy,
  and {MIMO} analysis,'' {\em IEEE Trans. Communications}, vol.~62,
  pp.~770--785, March 2014.

\bibitem{Yu19}
J.~Yu, W.~Gong, J.~Liu, L.~Chen, and K.~Wang, ``On efficient tree-based tag
  search in large-scale rfid systems,'' {\em IEEE/ACM Trans. Netw.}, vol.~27,
  pp.~42--55, Feb. 2019.

\bibitem{Luo19}
Z.~Luo, Q.~Zhang, Y.~Ma, M.~Singh, and F.~Adib, ``3d backscatter localization
  for fine-grained robotics,'' in {\em Proceedings of the 16th USENIX
  Conference on Networked Systems Design and Implementation}, NSDI'19,
  (Berkeley, CA, USA), pp.~765--781, USENIX Association, 2019.

\bibitem{Ma18}
Y.~Ma, Z.~Luo, C.~Steiger, G.~Traverso, and F.~Adib, ``Enabling deep-tissue
  networking for miniature medical devices,'' in {\em Proceedings of the 2018
  Conference of the ACM Special Interest Group on Data Communication}, SIGCOMM
  '18, (New York, NY, USA), pp.~417--431, ACM, 2018.

\bibitem{Liu13}
V.~Liu, A.~Parks, V.~Talla, S.~Gollakota, D.~Wetherall, and J.~R. Smith,
  ``Ambient backscatter: Wireless communication out of thin air,'' {\em SIGCOMM
  Comput. Commun. Rev.}, vol.~43, pp.~39--50, Aug. 2013.

\bibitem{Kellogg14}
B.~Kellogg, A.~Parks, S.~Gollakota, J.~R. Smith, and D.~Wetherall, ``Wi-fi
  backscatter: Internet connectivity for {RF}-powered devices,'' {\em SIGCOMM
  Comput. Commun. Rev.}, vol.~44, pp.~607--618, Aug. 2014.

\bibitem{Wang16}
G.~Wang, F.~Gao, R.~Fan, and C.~Tellambura, ``Ambient backscatter communication
  systems: Detection and performance analysis,'' {\em IEEE Trans.
  Communications}, vol.~64, pp.~4836--4846, Nov 2016.

\bibitem{Bharadia15}
D.~Bharadia, K.~R. Joshi, M.~Kotaru, and S.~Katti, ``{BackFi}: High throughput
  {W}i{F}i backscatter,'' {\em SIGCOMM Comput. Commun. Rev.}, vol.~45,
  pp.~283--296, Aug. 2015.

\bibitem{Kang17}
C.~{Kang}, W.~{Lee}, Y.~{You}, and H.~{Song}, ``Signal detection scheme in
  ambient backscatter system with multiple antennas,'' {\em IEEE Access},
  vol.~5, pp.~14543--14547, 2017.

\bibitem{Yang18B}
G.~Yang, Q.~Zhang, and Y.~Liang, ``Cooperative ambient backscatter
  communications for green internet-of-things,'' {\em IEEE Internet of Things
  Journal}, vol.~5, pp.~1116--1130, April 2018.

\bibitem{Yang19}
G.~{Yang}, D.~{Yuan}, Y.~{Liang}, R.~{Zhang}, and V.~C.~M. {Leung}, ``Optimal
  resource allocation in full-duplex ambient backscatter communication networks
  for wireless-powered {IoT},'' {\em IEEE Internet of Things Journal},
  pp.~1--1, 2019.

\bibitem{Yang18}
G.~Yang, Y.~Liang, R.~Zhang, and Y.~Pei, ``Modulation in the air: Backscatter
  communication over ambient {OFDM} carrier,'' {\em IEEE Trans.
  Communications}, vol.~66, pp.~1219--1233, March 2018.

\bibitem{Dahlman13}
E.~Dahlman, S.~Parkvall, and J.~Skold, {\em 4G: LTE/LTE-Advanced for Mobile
  Broadband, 2nd Edition}.
\newblock Academic Press, 2013.

\bibitem{Eizmendi14}
I.~Eizmendi, M.~Velez, D.~Gómez-Barquero, J.~Morgade, V.~Baena-Lecuyer,
  M.~Slimani, and J.~Zoellner, ``{DVB-T2}: The second generation of terrestrial
  digital video broadcasting system,'' {\em IEEE Trans. Broadcasting}, vol.~60,
  pp.~258--271, June 2014.

\bibitem{ProakisBook}
J.~Proakis, {\em Digital Communications}.
\newblock McGraw-Hill, fourth~ed., 2000.

\bibitem{Dempster77}
A.~P. Dempster, N.~M. Laird, and D.~B. Rubin, ``Maximum likelihood from
  incomplete data via the em algorithm,'' {\em J. of The Royal Statistical
  Society, Series B}, vol.~39, no.~1, pp.~1--38, 1977.

\bibitem{EM_Book}
G.~McLachlan and T.~Krishnan, {\em The EM Algorithm and Extensions}.
\newblock John Wiley \& Sons, 1997.

\bibitem{Qian17}
J.~Qian, F.~Gao, G.~Wang, S.~Jin, and H.~Zhu, ``Semi-coherent detection and
  performance analysis for ambient backscatter system,'' {\em IEEE Trans.
  Communications}, vol.~65, pp.~5266--5279, Dec 2017.

\bibitem{ChoBook}
Y.~S. Cho, J.~Kim, W.~Y. Yang, and C.~G. Kang, {\em MIMO-OFDM Wireless
  Communications with MATLAB}.
\newblock Wiley-IEEE Press, 2010.

\bibitem{ChoiJBook}
J.~Choi, {\em Adaptive and Iterative Signal Processing in Communications}.
\newblock Cambridge University Press, 2006.

\bibitem{Scharf}
L.~L. Scharf, {\em Statistical Signal Processing: Detection, Estimation, and
  Time Series Analysis}.
\newblock Addison-Wesley, 1991.

\bibitem{Chiani02}
M.~Chiani and D.~Dardari, ``Improved exponential bounds and approximation for
  the {Q}-function with application to average error probability computation,''
  in {\em Global Telecommunications Conference, 2002. GLOBECOM '02. IEEE},
  vol.~2, pp.~1399--1402 vol.2, Nov 2002.

\bibitem{SimonBook00}
M.~K. Simon and M.~Alouini, {\em Digital Communication over Fading Channels: A
  Unified Approach to Performance Analysis}.
\newblock John Willey, 2000.

\end{thebibliography}
